\documentclass{kapedbk}
\usepackage{graphicx,amsmath,amsfonts,edbkps}% Calls postscript fonts
  \normallatexbib

\begin{document}

\articletitle{Electron Dephasing in Mesoscopic Metal Wires}

\author{Norman O. Birge and F. Pierre\thanks{Permanent address: Laboratoire de Photonique et de Nanostructures (LPN)-CNRS, route de
Nozay, 91460 Marcoussis, France.}}

\affil{Department of Physics and Astronomy, Michigan State University, East Lansing, MI 48824-2320} \email{birge@pa.msu.edu}

\begin{abstract}
The low-temperature behavior of the electron phase coherence time, $\tau_{\phi}$, in mesoscopic metal wires has been a subject of controversy recently. Whereas theory predicts that
$\tau_{\phi}(T)$ in narrow wires should increase as $T^{-2/3}$ as the temperature $T$ is lowered, many samples exhibit a saturation of $\tau_{\phi}$ below about 1~K. We review here the
experiments we have performed recently to address this issue. In particular we emphasize that in sufficiently pure Ag and Au samples we observe no saturation of $\tau_{\phi}$ down to
our base temperature of 40~mK. In addition, the measured magnitude of $\tau_{\phi}$ is in excellent quantitative agreement with the prediction of the perturbative theory of Altshuler,
Aronov and Khmelnitskii. We discuss possible explanations why saturation of $\tau_{\phi}$ is observed in many other samples measured in our laboratory and elsewhere, and answer the
criticisms raised recently by Mohanty and Webb regarding our work.
\end{abstract}


\section{Introduction}

The Fermi liquid theory is at the root of the description of electronic properties of metals. It states that, despite the long range Coulomb interaction, the low-energy excitations
(quasiparticles) of a real metal have properties similar to those of a degenerate noninteracting electron gas. A key of the Fermi liquid theory is that as the temperature is reduced
scattering of quasiparticles becomes very rare due to phase space constraints by the Pauli exclusion principle. As a result the broadening of quasiparticle states due to interactions
becomes smaller. However the proof of Fermi liquid theory relies on momentum conservation and hence is valid only in perfectly crystalline metals with translational symmetry. In
disordered metals, such as metallic thin films deposited in e-gun or thermal evaporators, interactions between quasiparticles are stronger and energy dependent due to the reduced
mobility of electrons. Hence there is great interest in knowing whether Fermi liquid theory is still valid in disordered metals at arbitrary low temperatures. In that vein, measuring
the average lifetime of quasiparticle states $\tau_\phi$, also called the phase coherence time, provides a very powerful tool. Indeed, the description of low energy excitations by
quasiparticles holds only as long as the energy broadening $\hbar /\tau_\phi(T)$ of the quasiparticles states remains small compared to their average energy $k_{B}T$. Moreover, for
practical reasons, understanding the inelastic mechanisms which dominate the phase coherence time is crucial in the field of "mesoscopic physics" since many of the phenomena specific
to this field rely on quantum coherent transport \cite{meso}. Processes that destroy electron phase coherence, such as electron-electron and electron-phonon scattering, limit the
observability of most quantum-coherent mesoscopic phenomena. Hence understanding the sources of decoherence is essential to designing experiments or devices that rely on
quantum-coherent transport.

In the early 70's, Schmid showed that the electron-phonon and electron-electron scattering rates could be much larger in disordered metals than in perfectly crystalline metals
\cite{Schmid}. Altshuler, Aronov and coworkers then calculated the effect of electron-electron interactions on the quasiparticle lifetime and phase coherence in disordered metals
\cite{AAreview}. At temperatures below about 1~K, electron-phonon scattering occurs rarely, hence the dominant process limiting electron phase coherence (in the absence of magnetic
impurities) is electron-electron scattering.  The temperature dependence of the phase coherence time depends on the sample dimensionality; in quasi-1D wires it is predicted to diverge
with decreasing temperature as $\tau_\phi \propto T^{-2/3}$ \cite{AAK}.  That prediction was verified by experiments on Al wires (at temperatures down to 2 K) in 1986 \cite{Wind} and
on Au wires (down to 100~mK) in 1993 \cite{Echternach}.

It came as a surprise in 1997 when Mohanty, Jariwala, and Webb (MJW) \cite{MJW} reported measurements of $\tau_\phi(T)$ in six Au wires showing that, in contrast to theoretical
expectations, $\tau_{\phi}$ saturated at low temperature.  MJW speculated that saturation of $\tau_{\phi}$ at low temperature was an intrinsic, universal property of disordered metal
wires.  The same year, Pothier \textit{et al.} showed that information about electron-electron interactions could also be deduced from measurements of the quasiparticle energy
distribution function in a wire driven far from equilibrium by an applied voltage \cite{PRLrelax}. In their first experiments on Cu wires, Pothier \textit{et al.} observed that the
rate of energy exchange between quasiparticles was considerably larger than predicted by the theory of Altshuler and Aronov. Together, the experiments by MJW and by Pothier \textit{et
al.} suggested either that the standard theoretical picture of electron-electron interactions was incomplete, or that there was an extrinsic yet ubiquitous external factor playing an
important role in both experiments.

Since 1997, we and our collaborators at Saclay have performed measurements of $\tau_\phi(T)$ and of energy exchange on many samples of Au, Ag, and Cu \cite{Gougam, PierreJLTP,
PierrePesc, PierrePhD, AnthoreCu, AnthorePRL, PierreAB, PierrePRB}. A complete description of the $\tau_\phi(T)$ measurements was recently published \cite{PierrePRB}.  In this article
we summarize our main conclusions, present some additional data on Aharonov-Bohm conductance oscillations that did not appear in our earlier publications \cite{PierreAB, PierrePRB} and
answer the criticisms raised recently by Mohanty and Webb regarding our work \cite{MWrecent}.

\section{$\tau_{\phi}$ in pure Au and Ag samples}

We determine the phase coherence time $\tau_{\phi}$ in our wires by measuring the magnetoresistance in a magnetic field applied transverse to the wire, and fitting the data to the
expression from weak localization theory, with the phase coherence length $L_\phi$ as a fit parameter.  ($L_{\phi}$ is related to $\tau_{\phi}$ through the relation $L_{\phi} = \sqrt{D
\tau_{\phi}}$, where $D$ is the electron diffusion constant.)  The most reliable results are obtained if the wire is much longer than $L_\phi$ to limit the influence of measurement
leads and to ensure that the random but reproducible "Universal Conductance Fluctuations" (UCF) are much smaller than the weak localization contribution to the magnetoconductance.
Experimental details and the theoretical fitting function are provided in our earlier publications \cite{Gougam, PierrePRB}.

\begin{figure}[ht]\centering
\vskip .2in
\includegraphics[width=3.2in]{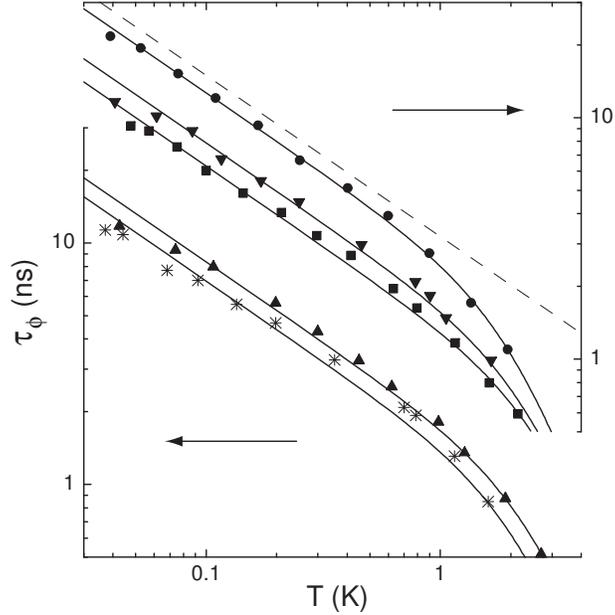} \caption{Phase coherence time vs temperature in 4 Ag samples (full symbols) and 1 Au sample ($\ast$),
all fabricated from source materials of 99.9999\% purity. Continuous lines are fits of the data to Eq.~(\ref{fitee-eph}). For clarity, the graph has been split in two part, shifted
vertically one with respect to the other. The quantitative prediction for electron-electron interactions in sample Ag(6N)c ($\bullet$) is shown as a dashed line. Taken from
\cite{PierrePRB}.} \label{FigAg6N}
\end{figure}

Our most important results are depicted in Fig.~\ref{FigAg6N}.  Here we show $\tau_{\phi}(T)$ for five samples fabricated from Ag or Au source material of 99.9999\% purity.  In these
samples, $\tau_{\phi}$ does not saturate, but rather continues to increase down to 40~mK -- the base temperature of our dilution refrigerator.  The solid lines in the figure are fits
to the function:
\begin{equation}
\tau_{\phi}^{-1}=AT^{2/3}+BT^{3}, \label{fitee-eph}
\end{equation}
where the first term describes dephasing by electron-electron scattering and the second term by electron-phonon scattering. Without further analysis, these data show that saturation of
$\tau_{\phi}$ is not a universal property of disordered metals, at least not at temperatures above 40~mK.  Not only do the data obey the $T^{-2/3}$ temperature dependence at low
temperature, but in addition the magnitude of the prefactor \textit{A} in Eq.~(\ref{fitee-eph}) is in excellent agreement with the theoretical prediction \cite{AAK, AleinerWav}. (See
\cite{PierrePRB} for a detailed quantitative comparison.)

A close inspection of Fig.~\ref{FigAg6N} shows that there are slight deviations from the $T^{-2/3}$ behavior of $\tau_{\phi}(T)$ at the lowest temperatures measured.  Hence one might
argue that $\tau_{\phi}$ does indeed saturate in these samples, but at temperatures below 40~mK.  Obviously, we can not prove that such a statement is incorrect.  We can, however,
compare these data with measurements made by MJW on similar samples.  For example, our samples Ag(6N)a and Au(6N) (see \cite{PierrePRB} for sample parameters) have similar geometrical
dimensions and diffusion constants as MJW's sample Au-3, and yet the maximum values of $\tau_{\phi}$ in the former two samples are 9 ns and 11 ns, respectively, but only 2 ns in MJW's
Au-3.  Our sample Ag(6N)c has a similar diffusion constant as MJW's Au-6, whereas its maximum value of $\tau_{\phi}$ is 22 ns compared with 3 ns for Au-6. From these comparisons it is
impossible to escape the conclusion that the saturation observed by MJW, at least on the above mentioned samples, is not a universal property of disordered metals, but is rather due to
some extrinsic factors.

Similar arguments have led us to reject a theoretical proposal by Golubev and Zaikin that there is always an intrinsic saturation of $\tau_{\phi}$ caused by zero-point fluctuations of
the electromagnetic field in the sample \cite{Zaikin}.  That theory was originally proposed to explain the data by MJW, which we now believe to be extrinsic. Furthermore, we have shown
in ref. \cite{PierrePRB} that the theory of Golubev and Zaikin is not consistent with the large body of $\tau_{\phi}$ data on metals now available in the literature and in particular
not even with only the measurements made by MJW.  A similar conclusion was reached by Gershenson, who compared the Golubev-Zaikin predictions with $\tau_{\phi}$ data in 2D metals and
semiconductors \cite{Gershenson}.

Note that the arguments presented above do not prove that there is no intrinsic saturation of $\tau_{\phi}$ at lower temperatures. The only way to address this question experimentally
is to measure $\tau_{\phi}$ in very pure samples at lower temperature. Unfortunately there have been few measurements of $\tau_{\phi}$ in pure samples at temperatures below 40~mK. The
nice measurements by Mueller \textit{et al.} of $\tau_{\phi}$ down to 20~mK in 2D Au films are unfortunately believed to be dominated by a small (sub-ppm) concentration of magnetic
impurities \cite{Mueller}. Recently there have been measurements of $\tau_{\phi}$ down to 20~mK in a two-dimensional electron gas using an innovative technique, namely a measurement of
Aharonov-Bohm conductance oscillations in a large array of connected rings \cite{Bouchiat}.  What was particularly valuable about this experiment is that several field harmonics were
measured, and $\tau_{\phi}$ was determined by ratios of amplitudes of adjacent harmonics.  The determination of $\tau_{\phi}$ then is very accurate because it does not depend on the
measurement leads connected to the rings. In this experiment, $\tau_{\phi}$ was found to obey the theoretically-predicted temperature dependence extremely accurately down to 20~mK.

Returning to the issue of the deviations from theory exhibited by the data shown in Fig.~\ref{FigAg6N} at very low temperature, we have shown previously that they could be accounted
for by residual magnetic impurities at concentrations of the order of 0.01~ppm, i.e. one impurity atom for every $10^8$ host atoms \cite{PierrePRB}.

\section{Possible explanations for saturation of $\tau_{\phi}$}

In spite of our strong belief that there is no experimental evidence for an intrinsic saturation of $\tau_{\phi}$, it is nonetheless a fact that an apparent saturation is observed in
many experiments, not only by MJW, but also by many other groups including ourselves.  Of the many samples we have measured, we observed saturation of $\tau_{\phi}$ in all (two) Ag
samples made from source material of 99.999\% (5N) purity rather than 99.9999\% (6N) purity, and also in all (six) Cu samples made from sources of both 5N and 6N purity
\cite{PierrePRB}. Two of the three Au samples we measured contained large concentrations of Fe impurities, which dominated their low temperature dephasing \cite{PierrePesc}.  A number
of possible explanations have been proposed to explain this saturation. One of the first explanations to appear was that external electromagnetic interference could cause electron
dephasing even without noticeably heating the electrons \cite{AGA}.  We can rule out this explanation for our own samples, because all of our samples were measured in the same cryostat
at Michigan State University, and yet we have observed saturation of $\tau_{\phi}$ in some samples but not in others of similar geometry and electrical resistance. Nevertheless, we can
not rule out the role of external interference in experiments performed by other groups; adequate filtering of the leads going into the cryostat is a constant concern for low
temperature experimenters.

A second proposal is that saturation of $\tau_{\phi}$ occurs due to interactions between the conduction electrons and two-level tunneling systems (TLS) in the metal (or perhaps even in
the oxide on the substrate) \cite{IFS, 2CK}.  We emphasize that such a result does not follow from the standard model of TLS \cite{AHVP}, which would lead to a dephasing rate
proportional to the temperature \cite{Black}.  Imry, Fukuyama, and Schwab found saturation of $\tau_{\phi}$ only if they assumed that the distribution of tunneling matrix elements had
an upper bound that was smaller than the temperature. (At lower temperatures they found $\tau_{\phi}(T) \propto T$, in agreement with the standard result.) It was pointed out later
\cite{AleinerTLS} that the TLS distribution proposed by Imry \textit{et al.} would lead to a large anomaly in the specific heat of the metal at very low temperature.  Such a specific
heat anomaly has never been observed in disordered metals, although data are only available down to 100~mK \cite{Graebner}. One could argue that the distribution of tunneling matrix
elements proposed in \cite{IFS} might be more applicable to polycrystalline metals than to amorphous metals \cite{ImryOvadyahu}. But measurements of \textit{1/f} resistance noise in
polycrystalline metals are consistent with the standard TLS distribution \cite{BirgeGolding}. A second proposal linking TLS to dephasing was based on the two-channel Kondo model
\cite{2CK}.  The proposal is that a subset of the TLS with nearly symmetric double-well potentials and strong coupling to the conduction electrons could cause $\tau_{\phi}$ to saturate
in the temperature range $\Delta < T < T_K$, where $\Delta$ is the tunneling matrix element and $T_K$ is the Kondo temperature.  There is strong theoretical evidence that the strong
coupling limit $\Delta < T_K$ of the two-channel Kondo model is inaccessible for real TLS in solids \cite{AleinerCont}, although this topic is still under investigation
\cite{BordaZawa}.

The third proposal is that saturation of $\tau_{\phi}$ is caused by magnetic impurities.  This is an old idea, dating back to the pioneering work of Hikami \textit{et al.}
\cite{Hikami}. Indeed, in the 1980's, many experimenters who observed saturation of $\tau_{\phi}$ attributed the behavior to small quantities of residual magnetic impurities
\cite{Pannetier, LinGiordano}.  The main reason why this topic is still under discussion is that MJW specifically rejected this explanation for their data on Au wires.  They based this
rejection on the fact that the most common magnetic impurity in Au is Fe, which causes a non-monotonic temperature dependence of $\tau_{\phi}(T)$ in the probed temperature range. In
the presence of Fe impurities, $\tau_{\phi}$ may exhibit a saturation at temperatures above 0.3 K (the Kondo temperature of Fe in Au), but it then "desaturates" (i.e. increases) at
lower temperatures \cite{MJW, PierrePesc, sami}.

\begin{figure}[ht]\centering
\vskip .2in
\includegraphics[width=3.1in]{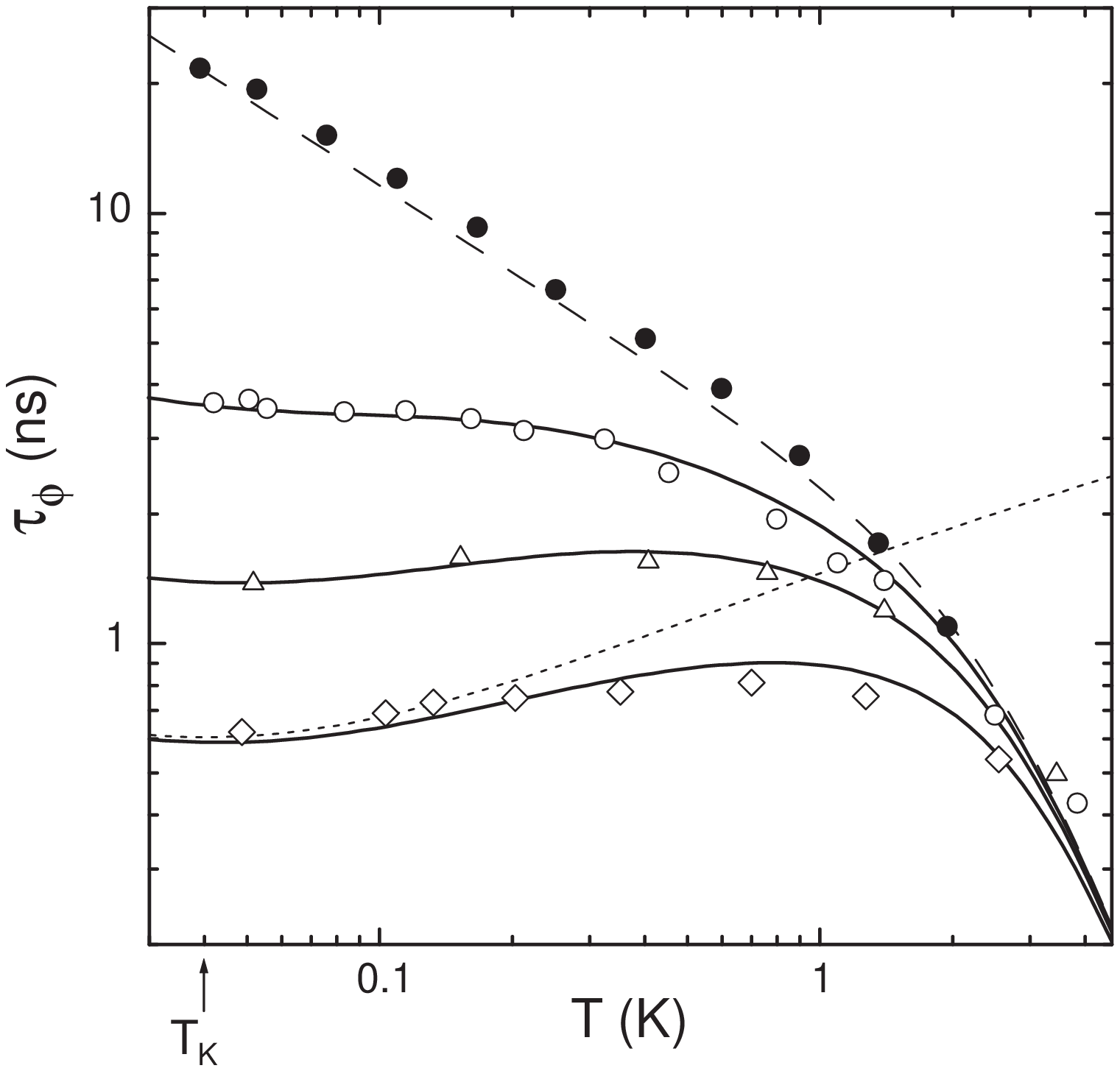} \caption{Phase coherence time as function of temperature in several silver wires. Sample Ag(6N)c ($\bullet$) is made of the purest
silver source. Samples Ag(5N)b ($\circ$), Ag(5N)c$_{\mathrm{Mn}0.3}$ ($\triangle$) and Ag(5N)d$_{\mathrm{Mn}1}$ ($\diamond$) were evaporated simultaneously using our 5N silver source.
Afterward, 0.3~ppm and 1~ppm of manganese was added by ion implantation respectively in samples Ag(5N)c$_{\mathrm{Mn}0.3}$ and Ag(5N)d$_{\mathrm{Mn}1}$. The presence of very dilute
manganese atoms, a magnetic impurity of Kondo temperature $T_{K}=40$~mK, reduces $\tau_{\phi}$ leading to an apparent ``saturation'' at low temperature. Continuous lines are fits of
$\tau_{\phi}(T)$ taking into account the contributions of electron-electron and electron-phonon interactions (dashed line) and spin flip collisions using the concentration
$c_{\mathrm{mag}}$ of magnetic impurity as a fit parameter (dotted line is $\tau_{\mathrm{sf}}$ for $c_{\mathrm{mag}}=1$~ppm). Best fits are obtained using $c_{\mathrm{mag}}=0.13,$
0.39 and 0.96~ppm respectively for samples Ag(5N)b, Ag(5N)c$_{\mathrm{Mn}0.3}$ and Ag(5N)d$_{\mathrm{Mn} 1}$, in close agreement with the concentrations implanted and consistent with
the source material purity used.  Taken from \cite{PierrePRB}.} \label{FigAgMn}
\end{figure}

Although we agree with MJW that Fe impurities in Au do not lead to a saturation of $\tau_{\phi}$ below $0.3 \mathrm{K}$, we nonetheless disagree with ruling out magnetic impurities
altogether. Other magnetic ions in Au, such as Cr or Mn, have much lower Kondo temperatures \cite{Wohlleben}. To test whether magnetic impurities with low $T_K$ can lead to saturation
of $\tau_{\phi}$, we have measured $\tau_{\phi}$ in Ag samples intentionally doped with very small concentrations of Mn impurities.  Fig.~\ref{FigAgMn} shows our results
\cite{PierrePRB}. The top curve shows $\tau_{\phi}$ in one of the very pure (6N) Ag wires shown earlier, for comparision.  The second curve shows $\tau_{\phi}$ in a Ag wire of slightly
less purity (5N instead of 6N), and the two lower curves show $\tau_{\phi}$ for Ag wires implanted with 0.3~ppm and 1.0~ppm Mn impurities, respectively.  The implanted wires were
fabricated from 5N purity Ag (the 6N source was not available at the time of this experiment), and the bare Ag wire Ag5Nb already shows a saturation of $\tau_{\phi}$.  The crucial
point we demonstrate here is that by implanting Mn impurities we further reduce $\tau_{\phi}$ and that it is still nearly independent of temperature below $1 \mathrm{K}$.  To
understand these results, we have fit $\tau_{\phi}(T)$ to a sum of dephasing rates due to electron-phonon, electron-electron, and spin-flip scattering processes. The latter process,
shown by the dotted line for a Mn concentration of 1~ppm, is represented by the Nagaoka-Suhl approximation, which is expected to be valid for temperatures $T>T_K$.  The concentrations
of Mn impurities deduced from the fits are in close agreement with the nominal concentrations determined during the ion implantation process \cite{PierrePRB}. The fits show that the
opposite temperature dependences of the electron-electron and spin-flip scattering rates can lead to a nearly temperature-independent $\tau_{\phi}$ over a substantial temperature
range.  Once the concentration of Mn exceeds about 1~ppm, a noticeable dip in $\tau_{\phi}$ (i.e. a peak in the spin-flip scattering rate) is apparent. Note that for such low
concentration of magnetic impurities the Kondo contribution to the temperature dependence of the resistivity is invisible, hidden by the larger contribution of electron-electron
interactions \cite{PierrePRB}.

\section{Aharonov-Bohm experiments in large Cu rings}

The results of the previous section demonstrate that the presence of magnetic impurities with low Kondo temperature can lead to an observed saturation of $\tau_{\phi}$, if their
concentration is in the range 0.1 - 1~ppm.  But that does not mean that whenever one observes saturation of $\tau_{\phi}$, it must be due to magnetic impurities.  What one needs is a
method to measure directly the presence of magnetic impurities at very low concentrations \cite{magneticimpurities}.  Since $\tau_{\phi}$ is itself extremely sensitive to magnetic
impurities, a good way to detect minute quantities of impurities is to measure $\tau_{\phi}$ as a function of the applied magnetic field $B$. Once $B$ is large enough such that $g\mu_B
B >> k_B T$, the magnetic impurity spins are frozen into their ground states, and there is no longer dephasing by spin-flip scattering \cite{spindephasing}. $\tau_{\phi}$ should then
increase to the value determined solely by electron-electron scattering (in absence of other extrinsic dephasing mechanisms).

There are two methods from mesoscopic physics that allow one to measure $\tau_{\phi}$ in high field.  One is to measure the amplitude and characteristic field scale of the universal
conductance fluctuations (UCF) in a narrow wire, and the other is to measure the amplitude of Aharonov-Bohm (AB) conductance oscillations in a small ring.  Our experiments are in the
temperature regime $L_T < L_{\phi}$, where $L_T = \sqrt{\hbar D/k_B T}$ is the thermal length.  In that regime the UCF amplitude is proportional to $L_{\phi}^{1/2}$ (or
$\tau_{\phi}^{1/4}$), and the UCF characteristic magnetic field scale is proportional to $1/L_{\phi}$.  The amplitude of Aharonov-Bohm oscillations, on the other hand, varies
exponentially with the ratio of the ring circumference to $L_{\phi}$, hence in principle it can be much more sensitive than UCF to small changes in $L_{\phi}$. Moreover, because the AB
oscillations are much narrower in $B$ than UCF fluctuations, it is possible to measure the full dependence of the phase coherence time $\tau_{\phi}(B)$ with the applied magnetic field,
rather than only its high field value. We note that Benoit \textit{et al.} demonstrated many years ago, in a pioneering experiment, that AB oscillations are strongly reduced at low
field in presence of a relatively large concentration (40 and 120~ppm) of magnetic impurities but increase dramatically when the magnetic field is large enough to freeze the spins
\cite{Benoit}. In contrast, we demonstrated that this known effect can be used to detect extremely dilute magnetic impurities \cite{PierreAB}.

\begin{figure}[ht]\centering
\vskip .2in
\includegraphics[width=4.1in]{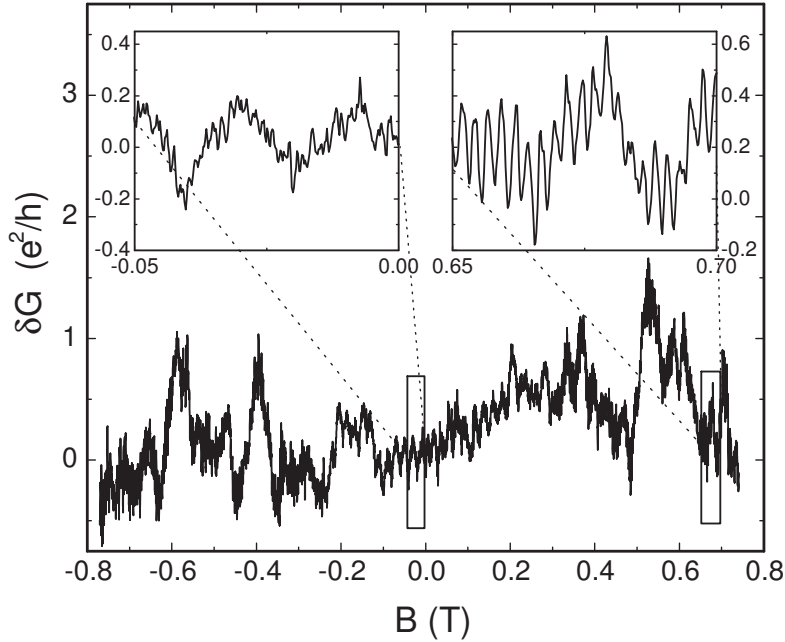}
\caption{Measured conductance changes of the ring in sample Cu4, in units of $e^2/h$, as a function of magnetic field at a temperature $T=40$~mK. The narrow Aharonov-Bohm oscillations
($\Delta B\simeq2.5$~mT) are superimposed on the larger and much broader universal conductance fluctuations. Left inset: blowup of the data near zero field. The AB oscillations are
hardly visible. Right inset: blowup of the data at large magnetic field. The AB oscillations are much larger.} \label{FigAB40mK}
\end{figure}

We chose to perform the AB experiment in Cu rings, because we always observed saturation of $\tau_{\phi}$ in Cu, even in samples made from 99.9999\% purity source material. To optimize
the sensitivity of the AB amplitude to changes in $\tau_{\phi}$, one should make the ring circumference as large as possible without decreasing the AB signal below the noise floor of
the experiment (a rule of thumb is to make the circumference comparable to $L_{\phi}$).  Fig.~\ref{FigAB40mK} shows the conductance vs. magnetic field of a Cu ring of diameter
$1.5~\mathrm{\mu m}$ at $T = 40~\mathrm{mK}$. (Raw data from the same sample at $T = 100~\mathrm{mK}$ were published in \cite{PierreAB}.)  The main part of the figure shows
predominantly UCF, as the AB oscillations occur on a field scale too narrow to discern on this scale. The insets show blow-ups of the field regions from -0.05 to 0 T, and from 0.65 to
0.7 T. In the first region, the AB oscillations are essentially invisible, and one sees only UCF. In the second region, large AB oscillations are clearly present. Their periodicity is
$\approx 2.4~\mathrm{mT}$, consistent with the nominal ring diameter of $1.5~\mathrm{\mu m}$.  Returning to the main part of the figure, it is now apparent that the AB oscillations
(and maybe also the UCF) are small near zero field, and grow considerably once B exceeds about 0.2 - 0.3 T in absolute value. Similar data at 100~mK show that this characteristic field
scale is proportional to the temperature (see Fig. 4 in \cite{PierreAB}).  In addition, the data of AB oscillation amplitude as a function of magnetic field can be fit quantitatively
with models describing the freezing out of the spin-flip scattering process by the magnetic field \cite{PierreAB, PierrePRB, VavilovGlazman}. At high field, the data are consistent
with the values of $\tau_{\phi}$ predicted by the theory of electron-electron scattering \cite{AAK}. Together, these observations confirm the hypothesis that the small value of
$\tau_{\phi}$ extracted from the low-field magnetoresistance in Cu samples is due to spin-flip scattering by magnetic impurities, most likely in the Cu oxide at the surface of our
wires \cite{vanHaesendonck}.

In a recent paper, Mohanty and Webb (MW) \cite{MWrecent} criticized our work on the AB effect, stating that "it is not clear as to why the peak-to-peak amplitude of the oscillations
even at the highest field is more than an order of magnitude smaller than the usual value of $e^2/h$ (see, for instance, Fig. 10b of Ref. \cite{WashburnWebb})."  Actually, the reason
for this apparent discrepancy is explained very clearly in the review article by Washburn and Webb \cite{WashburnWebb}; it is due to the relative size of the ring and the thermal
length $L_T = \sqrt{\hbar D/k_B T}$.  We list these parameters in Table~\ref{ABexperiments} for our large Cu ring and for the larger of the two Au rings studied by Webb and coworkers
in the mid-80's \cite{Webb, Washburn, WashburnWebb}. Our experiments on the Cu ring were performed in the limit $L_T < \pi r$ (or equivalently $k_B T > E_c$, where $E_c = \hbar D/(\pi
r)^2$ is the correlation energy), hence the AB oscillation amplitude is reduced by a factor $L_T/\pi r$ relative to the zero-temperature result.  At $T = 100~\mathrm{mK}$, the
peak-to-peak amplitude of the oscillations shown in Fig. 3 of our paper \cite{PierreAB} is about $0.2 e^2/h$.  Fig.~\ref{FigAB40mK} above shows that the AB amplitude is approximately
twice as large at 40~mK. (That could also be deduced from Fig. 4 in \cite{PierreAB}.)  From Table I, we expect the AB amplitude to reach the zero-temperature limit $e^2/h$ only at very
low temperature -- about 7~mK, when $L_T \approx \pi r$.  In contrast, the Au ring from the earlier work was in the limit $L_T > \pi r$ for temperatures below about 30~mK. The
crossover from $L_T > \pi r$ to $L_T < \pi r$ is shown in Fig. 12 of \cite{WashburnWebb} as a function of temperature for a small Au ring and as a function of voltage for the large
ring.  (In the latter case, $L_T$ is replaced by $L_V = \sqrt{\hbar D/eV}$.)

\begin{table}[ht]
\caption{Comparison of Aharonov-Bohm experiments: sample size, thermal length, dephasing length.  The values of $L_T$ and $L_{\phi}$ given in the table are evaluated at the
temperatures shown (different for each experiment), and at high magnetic field for the Cu sample.}

\begin{tabular*}{\textwidth}{@{\extracolsep{\fill}}lccccccc}
%\begin{tabular}{|l|ccccccc|}
\sphline
          &       &  $D$  & $T$  & $L_T$    & $\pi r$  & $L_{\phi}$ &         \\
material  & ref.  &($\mathrm{cm^2/s}$)& (mK) & ($\mathrm{\mu m}$)&($\mathrm{\mu m}$)&($\mathrm{\mu m}$) & regime  \\
\sphline
Cu  &  \cite{PierreAB}  &  52   &  40  &  1.0  & 2.4  & 7.0  & $L_T < \pi r < L_{\phi}$  \\
Au  &  \cite{Webb}      &  68   &  10  &  2.3  & 1.3  & ~1.7   & $\pi r < L_T \approx L_{\phi}$ \\
Ag  &  \cite{ChandraAB} &  100   &  1.7  &  0.25  & 1.6  & 0.9  & $L_T < L_{\phi} < \pi r$ \\
\sphline
\end{tabular*} \label{ABexperiments}
\end{table}

Similar considerations relating to the relative sizes of the three length scales -- $L_{\phi}$, $L_T$, and $\pi r$ -- explain the other observations regarding AB oscillations made by
MW \cite{MWrecent}. Their comment that AB oscillations in the early samples displayed no field dependence is consistent with the fact that $L_{\phi} > \pi r$ in those samples already
at B=0 \cite{WebbComment}. MW also comment that the temperature dependence of the AB amplitude in the earlier work is governed primarily by $L_T$ rather than by $L_{\phi}$. Again, that
is true in the regime $L_T < \pi r < L_{\phi}$, but is no longer the case when $L_{\phi} < \pi r$. The latter situation was explored by Chandrasekhar \textit{et al.}, who measured AB
oscillations in Al and Ag rings at much higher temperature.  Those workers observed that the AB oscillation amplitude decreased with temperature faster than $T^{-1}$, presumably due to
both the $L_T \propto T^{-1/2}$ prefactor and the T-dependence of $L_{\phi}$.

\section{Conclusions}

The most important conclusion from this work is that there is as yet no experimental evidence that the saturation of $\tau_{\phi}(T)$ often observed at low temperature is an intrinsic
property of disordered metals.  Rather, just the opposite is true.  In very pure Ag and Au samples, the low-temperature behavior of $\tau_{\phi}(T)$ is quantitatively consistent with
the theoretical predictions of Altshuler, Aronov, and Khmelnitskii \cite{AAK}.

We caution that our work does not answer definitively the question as to why a saturation of $\tau_{\phi}$ is often observed in mesoscopic metal samples.  In our own samples that
exhibit such a saturation, we have always been able to attribute it to the presence of dilute magnetic impurities.  That is also true in the case of experiments on energy exchange
performed at Saclay \cite{AnthorePRL} in parallel with our work.  But our results do not imply that the magnetic impurity explanation is universal.  In their recent paper
\cite{MWrecent}, Mohanty and Webb present data on Universal Conductance Fluctuations (UCF) up to high magnetic field, which they interpret as showing that the saturation of $L_{\phi}$
in those two samples is not due to magnetic impurities.  As discussed earlier, we believe that measurements of UCF are not the best way to determine $L_{\phi}$, because UCF are not
very sensitive to $L_{\phi}$. (Indeed, even putting aside questions about prefactors, the \textit{relative} size of $L_{\phi}$ in the two samples measured in \cite{MWrecent} is
opposite when $L_{\phi}$ is extracted from the UCF amplitude or from the UCF magnetic field scale.  See Figs. 3 and 4 in \cite{MWrecent}.) But even if the interpretation in
\cite{MWrecent} is correct, those experiments do \textit{not} justify the conclusion of that paper that saturation of $\tau_{\phi}$ is intrinsic.

Experimental measurements of $\tau_{\phi}$ have also been carried out in amorphous metallic alloys \cite{JJLin}.  It is our belief that control of sample purity in those systems is
considerably more difficult than in the noble metals, and we have already seen that even minute concentrations of magnetic impurities in the latter can lead to a drastic reduction of
$\tau_{\phi}$ relative to its intrinsic value.  On the other hand, the amorphous alloys are in the limit $k_F l_e \approx 1$, rather than $k_F l_e >> 1$ where the perturbative theories
\cite{AAK} are most likely to be valid ($k_F$ is the Fermi wavevector and $l_e$ is the mean free path).  Hence they represent a potentially fruitful area for further study.  At the
present time, however, we are hesitant to draw conclusions about "intrinsic" saturation of $\tau_{\phi}$ from that work \cite{LinBird}.

This work was supported by NSF grants DMR-9801841 and 0104178, and by the Keck Microfabrication Facility supported by NSF DMR-9809688.  We thank our collaborators: A. Anthore, M.
Devoret, D. Esteve, A. Gougam, S. Gu\'{e}ron, and H. Pothier.  We have also enjoyed interesting discussion with many people, including I.~Aleiner, B.L.~Altshuler, V.I.~Fal'ko,
L.I.~Glazman, M.G.~Vavilov and A.D. Zaikin.


\begin{chapthebibliography}{99}
\bibitem{magneticimpurities} At concentrations larger than a few ppm, the presence of magnetic impurities in mesoscopic wires can be detected from their logarithmic contribution to the
temperature-dependence of the resistivity.
\bibitem{spindephasing} Although frozen spins do not suppress UCF and the $h/e$ Aharonov-Bohm effect, they do suppress the weak localization contribution to the conductivity.  This
subtlety is discussed in \cite{VavilovGlazman}.
\end{chapthebibliography}
\end{document}